# Development of a pipeline CAMAC controller with PC/104-Plus single board computer

Y. Yasu, E. Inoue, H. Fujii, Y. Igarashi, M. Ikeno, M. Tanaka, K. Nakayoshi and H. Kodama
*High Energy Accelerator Research Organization (KEK), Tsukuba, Japan*

S. Harada
*TOYO corporation, Tokyo, Japan*

H. Kyoo
*Fird corporation, Tokyo, Japan*

A pipeline CAMAC controller with PC/104-Plus single board computer has been developed. The architecture of the pipeline CAMAC controller is completely different from that of traditional CAMAC controller. The pipeline CAMAC controller adopted truly pipelined command and data paths from/to computer memory to/from CAMAC. The pipeline method enabled the controller maximum speed of CAMAC, approximately up to 3 MB/sec with 24-bit data. The controller also has a DAQ function such as event numbering for co-existence with other equipment such as VME. Not only the functionality and the performance but also the pipeline architecture and the design concept are described.

## 1. INTRODUCTION

The Pipeline CAMAC controller occupies 2 slots (stations $24^{th}$ and $25^{th}$) in a CAMAC crate [1]. In Figure 1, there are the controller in a CAMAC crate and a PC connected to the controller via Fast Ethernet. The controller consists of a PC/104-Plus-based single board computer, PCI interface (PCI control) and CAMAC interface (CAMAC control). Thus, the controller can be used as an intelligent controller.

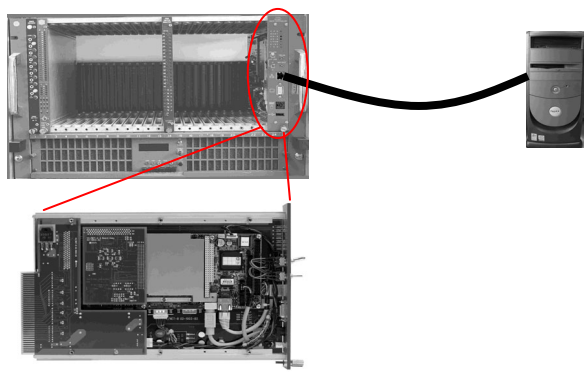

Figure 1: Pipeline CAMAC controller.

The PC/104-Plus is a standard PCI specification for embedded systems [2]. The board computer with low power consumption includes a flash disk with IDE interface, Fast Ethernet, USB and so on. Many PC/104-Plus based single board computers are available in the industry.

The PCI and CAMAC interfaces consist of an ALTERA FPGA, respectively [3]. Those VHDL codes for the interfaces have been also developed. The PCI control is originally designed for the pipeline CAMAC controller while the PCI cores available from commercial are not convenient from the pipeline point of view. The PCI control is also designed in general purpose. The communication protocol between the PCI and the CAMAC is well defined. Thus, the PCI control is usable for other I/O devices.

The controller has a DAQ function such as event numbering. The signal handling of trigger-input and busy-out are implemented while the event count can be read out. This feature is important to co-exist with other equipment such as VME which already implements timing & trigger control signaling.

From application point of view, the pipeline CAMAC controller will make the traditional system based on CAMAC powerful because it is an intelligent controller including CPU with Fast Ethernet and realizes maximum speed of CAMAC access.

## 2. FUNCTIONALITY

The pipeline CAMAC controller is located into $24^{th}$ and $25^{th}$ slots of CAMAC crate. It mainly consists of PC/104-Plus Single Board Computer, PCI control and CAMAC control.

The board computer adopted is Advantech PCM-9370, 3.5 inch (145 x 102 mm) Transmeta Crusoe 500 MHz processor single board computer including TM5400 processor, 320 MB memory, two IDE UltraDMA33 with 512 MB flash disk, LCD/CRT controller, 10/100 Mbps Ethernet controller, two 1.1 compliant USB ports, connector for keyboard, PS/2 mouse and so son [4]. The power consumption is typically 10.7W [4]. Those devices are available at the front panel of the controller as in Figure 2.

On the front panel, there are several indicators and switches. "Busy" indicates CAMAC Busy signal. Light "NO-Q" and light "NO-X" indicate that the CAMAC access does not response Q and X. Light "L-SUM" means there is at least a CAMAC module issuing Look-At-Me in the CAMAC crate. When "Online/Offline" switch is toggled to "Online", CAMAC is accessible. CAMAC Z and C signal can be manually issued by using "Z / C" switch. "TRG-IN" and "BSY-OUT" are NIM signals with LEMO connectors for trigger input and busy out, respectively.

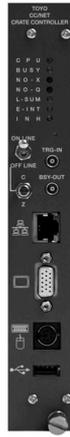

Figure 2: Front Panel of Pipeline CAMAC controller.

In the CAMAC control, there are a CAMAC Executor, a CAMAC LAM Handler (CAMAC Interrupter), a DAQ Executor and a DAQ Trigger Handler (DAQ Interrupter). Figure 3 shows the architecture of CAMAC control. The Executors accept a packet including at least a command frame or more from CPU via PCI, execute them and then send reply frames to CPU. The frame is a unit of CAMAC access. The pipeline CAMAC controller can execute CAMAC command with small overhead continuously by getting the command frames from a computer memory and then sending the reply frames to the computer memory, via PCI bus. The command FIFO and the reply FIFO have the size of 256 frames each. The Interrupters generate a packet including a reply frame and then send to CPU. When LAM or Trigger interrupt occurs during the Executor processes a packet, the generation of interrupt reply frame will be postponed. After the packet is processed, the interrupt reply frame will be sent.

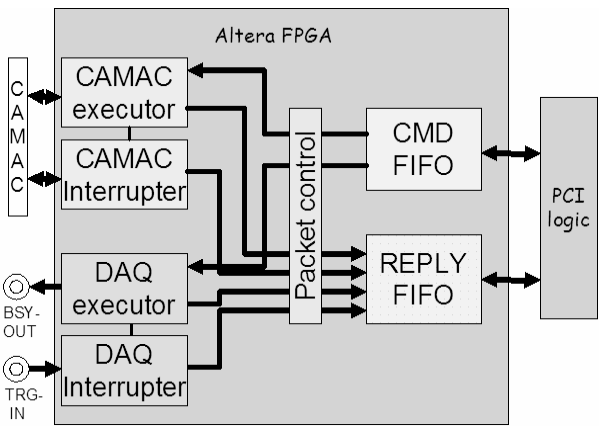

Figure 3: Architecture of CAMAC control.

The command and reply frames have 64-bit fixed-length. It consists of an 8-bit header and a 56-bit payload. The payload for basic CAMAC command frame contains CAMAC station number N, CAMAC sub-address A, CAMAC function F and data to be written if necessary.

That for the reply frame additionally contains CAMAC status such as Q & X and read data if the CAMAC function is 'read'. That for CAMAC LAM includes 24-bit LAM pattern and that for DAQ Trigger includes 32-bit event count. DAQ Executor can clear Busy-out signal while the signal disables next event trigger. Figure 4 and Figure 5 show the frame format of CAMAC function and DAQ function, respectively.

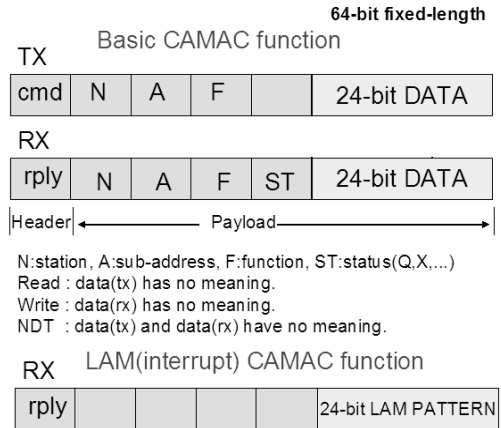

Figure 4: Frame format of CAMAC function

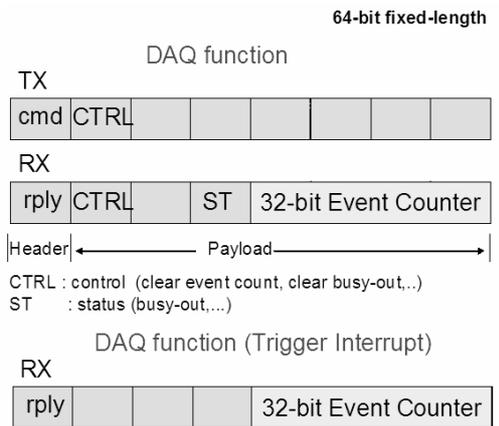

Figure 5: Frame format of DAQ function

The architecture of PCI control is shown in Figure 6. There are Tx/Rx FIFO, Tx/Rx engines and PCI multiplexer for Tx/Rx. Tx and Rx engines work independently while PCI multiplexer manages the PCI usage for Tx and Rx engines. Tx/Rx FIFOs have the depth of 256 with 32-bit width each. Tx/Rx engines have two modes, master and slave. The slave means PCI slave which the PCI interface becomes. In master mode, the PCI interface can transfer data from/to memory in computer without CPU intervention. Thus, master mode not only reduces CPU usage but also realizes fast transfer in comparison with the slave mode. From the programming point of view, the programmed I/O uses the PCI slave mode while Block I/O (DMA) uses the PCI master mode.

On one hand, the pipeline CAMAC controller sends multiple command frames, which include N, A, F and data if necessary, via Tx I/O registers. On the other hand, the controller receives multiple reply frames, which also includes N, A, F, status (Q and X) and data if necessary, via Rx I/O registers. The operations to Tx and Rx are done concurrently, independently, or asynchronously. If DMA is used in both sides, the throughput reaches maximum speed. DMA function is provided on traditional CAMAC controller [5], but the function is limited such as Address-scan, Q-stop and so on. For the pipeline CAMAC controller, any operations to CAMAC can be executed in DMA mode. There is no limitation.

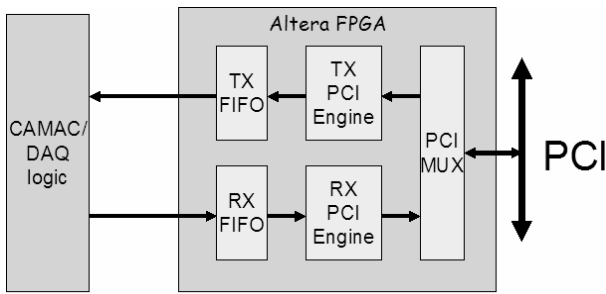

Figure 6: Architecture of PCI control

There are special CAMAC functions to the controller itself. For examples, CAMAC C and CAMAC Z are generated by those functions. All of the functions are shown in Table I. The fast cycle means S2 timing will be ignored in this mode. This is illegal from CAMAC protocol point of view, but there are lots of CAMAC modules usable in this mode.

Table 1: Special CAMAC Function

| Function | Description |
|---|---|
| N(25).A(0).F(24) | Clear CAMAC Inihibit |
| N(25).A(0).F(26) | Set CAMAC Inhibit |
| N(25).A(1).F(24) | Disable Interrupt to CPU |
| N(25).A(1).F(26) | Enable Interrupt to CPU |
| N(25).A(2).F(24) | Disable Fast Clcye |
| N(25).A(2).F(26) | Enable Fast Cycle |
| N(25).A(0).F(16) | Generate CAMAC C |
| N(25).A(0).F(17) | Generate CAMAC Z |
| N(25).A(1).F(0) | Read LAM Enable register |
| N(25).A(1).F(16) | Write LAM Enable register |

## 3.  OPERATION

The PCI I/O register map includes three kinds of registers. One is for Tx and another is for Rx. The other is system register. Those registers are located in PCI I/O space and the size of those registers is 32-bit. Table 2 shows the map.

In programmed I/O, data1 and data2 registers for Tx and Rx are used for sending command frames and receiving reply frames, respectively. FIFO count register contains number of data in long-word (4bytes). For an example, Rx FIFO count register is used for getting the number of data to be transferred as reply frames after sending command frames.

In block I/O, memory address register is used for pointing to address of kernel command/reply frame buffers and preset count register represents the number of frames in quad-word (8 bytes) to be transferred. After the transfer is initiated by setting DMA-start flag, the actual count register counts the number of frames in quad-word (8 bytes) transferred actually. Control register and status register contains control and status information related to interrupt and block transfer.

System register is used by system administrator.

Table 2: PCI I/O Register Map

| I/O offset address | Register name | Description |
|---|---|---|
| 00h | TxData1 | Tx data1 register |
| 04h | TxData2 | Tx data2 register |
| 08h | TxControl | Tx control register |
| 0Ch | TxStatus | Tx status register |
| 10h | TxAddress | Tx memory address register |
| 14h | TxPresetCount | Tx preset count register |
| 18h | TxActualCount | Tx actual count register |
| 1Ch | TxFifoCount | Tx FIFO count register |
| 20h | RxData1 | Rx data1 register |
| 24h | RxData2 | Rx data2 register |
| 28h | RxControl | Rx control register |
| 2Ch | RxStatus | Rx status register |
| 30h | RxAddress | Rx memory address register |
| 34h | RxPresetCount | Rx preset count register |
| 38h | RxActualCount | Rx actual count register |
| 3Ch | RxFifoCount | Rx FIFO count register |
| 40h | System | System register |

## 4.  SOFTWARE

The choice of a suitable Operating System (OS) for PC compatible hardware is dependent on the expected functionality. Since the software tools, like compilers and utilities are, Open Source Software / GNU products, a commodity Open Source OS, Linux is chosen. A Linux distribution kit tailored by KEK is implemented for the pipeline CAMAC controller.

The board computer has a flash disk. The flash disk has a limited life expectancy. The life expectancy is measured in the number of erase cycles. Linux frequently erases the directories of /var and /tmp. Thus, they are implemented as RAM disk.

The device driver and the library for the pipeline CAMAC controller have been developed. CAMAC and DAQ command frames are generated and then stored into a frame buffer by the frame generator. The command frames in a frame buffer are executed by Programmed I/O routines, Block I/O routines or the combined routines in same manner. The reply frames are stored into another frame buffer as result of the command frame executions. The reply frame includes CAMAC status and data. The extractors of the status and the data from the reply frames are provided. The wait routines for LAM interrupt and Trigger interrupt are also available [6], [7].

**TUGP004**

## 5. DESIGN CONCEPT

The pipeline CAMAC controller adopted truly pipelined command and data paths from/to computer memory to/from CAMAC. The pipeline method enabled the controller maximum speed of CAMAC access, approximately up to 3 MB/sec with 24-bit data. In the pipeline method, the multiple command frames and the multiple reply frames are processed asynchronously, independently or concurrently. This feature is different from that of the traditional CAMAC controller [8].

As described in previous section, DMA function on traditional CAMAC controller is limited such as Address-scan, Q-stop and so on. For the pipeline CAMAC controller, any operations to CAMAC can be executed in DMA mode. There is no limitation. Thus, the pipeline CAMAC controller is different from traditional CAMAC controller, from DMA point of view, too.

The design concept of pipeline CAMAC controller expands the feasibility of I/O method. The pipeline method is best for the speed up of I/O throughput. The method applied to the pipeline CAMAC controller is generally usable, namely, as the access method to I/O object controller, which includes not only CAMAC but also readout equipment such as Flash ADC as in Figure 7.

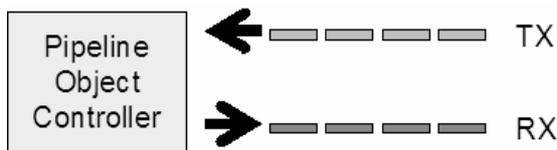

Figure 7: Access to I/O object

## 6. PERFORMANCE

CAMAC executor executes a CAMAC command frame in 1.04 usec. Figure 8 shows CAMAC timing. There are CAMAC Busy signal and CAMAC S1 signal in the vertical axis. The horizontal axis is timing in unit 1 usec. Each CAMAC cycle is continuously done in 1.04 usec. The speed was achieved by the DMA operation.

In fast mode which ignores CAMAC S2 timing signal, it executes the command frame in 0.72 usec.

## 7. CONCLUSIONS

A pipeline CAMAC controller with new architecture has been developed.

The CAMAC access speed reached up to maximum speed of CAMAC, approximately 3 MB/s with 24-bit data.

The pipeline method will be useful for not only CAMAC controller but also other controller, generally, I/O object controller.

### Acknowledgments

The authors would like to thank the people in KEK online and electronics group for their support.

### References

[1] IEEE, "IEEE Standard Modular Instrumentation and Digital Interface System (CAMAC) (Computer Automated Measurement and Control) IEEE Std 583-1975.
[2] PC/104 Embedded Consortium, PC/104-Plus Specification Version 1.2, August 2001
[3] Altera Corporation, http://www.altera.com/
[4] Advantech Co., Ltd., Board computer PCM-9370 http://www.advantech.co.jp/epc/SBC/biscuit/pcm9370.html
[5] KineticSystems Corp., VME to 3922 Interface w/DMA model 2917-Z1A & Model 2915-Z1A & PCI Interface to 3922 Model 2915-Z1A & Parallel Bus Crate Controller Model 3922-Z1B http://www.kscorp.com/www/products/camac.htm
[6] Home page of Parallel CAMAC project, http://www-online.kek.jp/~yasu/Parallel-CAMAC/
[7] Yet another page of Parallel CAMAC project in Japanese, http://www-online.kek.jp/~inoue/para-CAMAC/
[8] TOYO Corp., CAMAC Crate Controller type CC/7700 and Host Interface Board types CC/ISA and CC/PCI in Japanese http://www.toyo.co.jp/daq/index.html

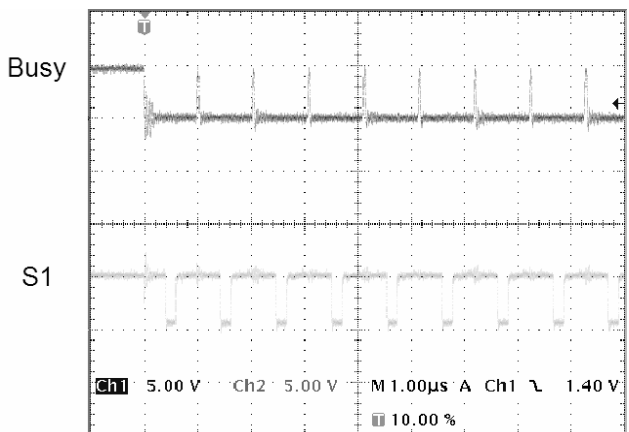

Figure 8: CAMAC timing